\documentclass[preprint,aps,showpacs]{revtex4}

\usepackage{graphicx}
\usepackage{dcolumn}
\usepackage{bm}
\usepackage{xspace}
\usepackage{natbib}
\usepackage{color}

\begin{document}

\preprint{}

\title{
Electronic structure of optimally doped pnictide Ba$_{0.6}$K$_{0.4}$Fe$_2$As$_2$: a comprehensive ARPES investigation
}

\author{
H. Ding,$^1$ K. Nakayama,$^2$ P. Richard,$^3$ S. Souma,$^3$ T. Sato,$^{2,4}$ T. Takahashi,$^{2,3}$ M. Neupane,$^5$ Y.-M. Xu,$^5$ Z.-H. Pan,$^5$
A.V. Federov,$^6$ Z. Wang,$^5$ X. Dai,$^1$ Z. Fang,$^1$ G.F. Chen,$^1$ J.L. Luo,$^1$ and N.L. Wang$^1$
}

\address{
 (1) Beijing National Laboratory for Condensed Matter Physics, and Institute of Physics, Chinese Academy of Sciences, Beijing 100190, China\\
 (2) Department of Physics, Tohoku University, Sendai 980-8578, Japan\\
 (3) WPI Research Center, Advanced Institute for Materials Research, Tohoku University, Sendai 980-8577, Japan\\
 (4) TRIP, Japan Science and Technology Agency (JST), Kawaguchi 332-0012, Japan\\
 (5) Department of Physics, Boston College, Chestnut Hill, MA 02467, USA\\
 (6) Advanced Light Source, Lawrence Berkeley National Laboratory, Berkeley, CA 94720
}

\date{\today}

\begin{abstract}
We have conducted a comprehensive angle-resolved photoemission study
on the normal state electronic structure of the Fe-based
superconductor Ba$_{0.6}$K$_{0.4}$Fe$_2$As$_2$. We have identified
four dispersive bands which cross the Fermi level and form two
hole-like Fermi surfaces around $\Gamma$ and two electron-like Fermi
surfaces around M.  There are two nearly nested Fermi surface pockets connected by an 
antiferromagnetic ($\pi$, $\pi$) wavevector. The observed Fermi surfaces show small $k_z$
dispersion and a total volume consistent with Luttinger theorem.
Compared to band structure calculations, the overall bandwidth is
reduced by a factor of 2. However, many low energy dispersions
display stronger mass renormalization by a factor of $\sim$ 4,
indicating possible orbital (energy) dependent correlation effects. Using an
effective tight banding model, we fitted the band structure and the
Fermi surfaces to obtain band parameters reliable for theoretical
modeling and calculations of the important physical quantities, such
as the specific heat coefficient.

\end{abstract}
\vspace{1.0cm}


\maketitle
\pagebreak

The recent discovery of superconductivity in iron pnictides has
opened a new route to high temperature superconductivity beyond the
cuprates \cite{Hosono}. It is now widely believed that the multiband
nature of this material is important, pertaining to the
superconducting instability in the doped compound and the
antiferromagnetic (AF) spin density wave (SDW) instability \cite
{neutron_Dai} in the parent compound. The knowledge of the
electronic band structure and Fermi surface (FS) topology is
critical to understanding the underlying physics. Most
first-principle band theory calculations, such as local density
approximation (LDA) \cite{LDA_Singh,LDA_Lu,LDA_Xu}, have predicted
that five bands of the Fe 3$d$ $t_{2g}$ complex cross the Fermi
level ($E_F$), and form three hole-like FSs centered at the zone
center ($\Gamma$) and two electron-like FSs centered at the zone
corner (M). However, there remain inconsistencies in the predicted
band structure \cite{review_Mazin}. More seriously, the optimal As
position calculated in LDA is quite different (more than 10\%) from
the experimental value \cite{review_Mazin}. On the experimental
side, angle-resolved photoemission spectroscopy (ARPES) studies \cite{HongEPL,
AdamNd, FengSr, AdamBK, ZhouBK, ShenP, Borisenko, Hassan} have
observed several dispersive bands and FSs, showing some
consistencies with LDA calculations. However, some major
discrepancies exist and a quantitative comparison is lacking. In
particular, there is no consensus regarding the band structure and
FS near the M point. In order to resolve these controversies and
thus obtain comprehensive knowledge of the electronic structure of
the iron pnictides, we have carried out a systemic ARPES study on an
optimally doped superconductor Ba$_{0.6}$K$_{0.4}$Fe$_2$As$_2$.

The high-quality single crystals of Ba$_{0.6}$K$_{0.4}$Fe$_2$As$_2$
($T_c$ = 37 K) used in this study were grown by the flux method
\cite{ChenGF}. These are the same samples used to determine the
FS-dependent nodeless superconducting gaps that close at the bulk
$T_c$ \cite{HongEPL}. Low-energy electron diffraction on a
mirror-like cleaved surface shows a sharp 1$\times$1 pattern without
any detectable reconstruction down to 20 K. All of these indicate
that the cleaved surface on this material retains a
bulk-representative electronic structure. High-resolution (4 - 20
meV) ARPES measurements were performed in the photoemission
laboratory of Tohoku University using a microwave-driven Helium
source ($h\nu$ = 21.218 eV), and several synchrotron beamlines in
the Synchrotron Radiation Center and the Advanced Light Source in
the US, and the Photon Factory in Japan, using various photon
energies ranging from 20 eV to 100 eV to selectively enhance
different band features of this multiband system. Samples were
cleaved $in$ $situ$ at low temperature (10 - 40 K) and measured at 7
- 150 K in a working vacuum better than 1$\times$10$^{-10}$ Torr.

We start with a wide energy spectrum (Fig.~1a) that includes shallow
core levels and the valence band, which can yield valuable
information on the valence and chemical environment of the
constituent elements. The strong double peaks at the binding
energies of 40.4 and 41.1 eV are from As 3$d_{5/2}$ and 3$d_{3/2}$,
the same as the 3$d$ core levels of As in bulk GaAs (40.4 and 41.1
eV) \cite{GaAs}. However, unlike GaAs (110) that has a surface
component in the As 3$d$ core levels,
Ba$_{0.6}$K$_{0.4}$Fe$_2$As$_2$ shows no clear evidence for a second
component of As 3$d$, suggesting no major surface modification
involving the As atoms. We have also identified several other core
levels, such as Fe 3$p$ (52.4, 53.0 eV), K 3$s$ (33.0 eV) and 3$p$
(17.8 eV), Ba 5$s$ (29.7 eV) and 5$p$ (14.2, 16.2 eV), and As 4$s$
(16.9 eV). In addition, there is a weak but well-defined peak at 12
eV, as shown in the inset of Fig.~1a, which has also been observed
in our measurements of the parent compounds BaFe$_2$As$_2$ and
SrFe$_2$As$_2$. A peak at a similar binding energy has been observed
in divalent iron compounds, such as FeO \cite{FeO}, which has been
attributed to a satellite state with the Fe 3$d^5$ configuration. A
valence satellite state, such as the well-known 11 eV Cu 3$d^8$
satellite observed in many cuprate superconductors, has been
attributed to strong correlation effects. The observation of the 12
eV satellite peak suggests the importance of the local electronic
correlations at the Fe sites in the pnictides.

The valence band shows a strong photon energy dependence. Comparing
energy distribution curves (EDCs) measured at 100 eV (enhancing the
intensity of Fe 3$d$) and the ones measured at 21.2 eV (enhancing
the intensity of As 4$p$), one can conclude that the strong peak
within 1 eV from $E_F$ is mostly from Fe 3$d$ orbitals, and the
states behind it are mostly from As 4$p$, consistent with the LDA
calculations \cite{LDA122_Lu}. The valence band intensity undergoes
a strong variation, or a resonance, when the photon energy is
scanned through the Fe 3$p$ absorption edge ($\sim$ 56 eV), as seen
in Fig.~1b. Following a common practice in photoemission \cite{FeO},
we plot in Fig.~1b the difference between EDCs measured at 56 eV (at
resonance) and 52 eV (below resonance), which corresponds mostly to
Fe 3$d$ states since the intensity of other orbitals is not expected
to change drastically over this narrow photon energy window \cite{footnote}. The
difference curve shows a sharp peak at $E_F$ which corresponds to
the coherent Fe 3$d$ orbitals, and a broad peak centered at $\sim$ 7
eV which can be regarded as the incoherent part of Fe 3$d$ states.
The assignment of coherent and incoherent components is also
supported by the observation in Fig.~1c of the ``anti-resonance''
profile of the coherent part and the Fano-like resonance profile for
the incoherent part at 7 eV and 12 eV (satellite peak) due to
super-Coster-Kronig Fe 3$p$-3$d$ Auger transition, similar to what
has been observation in FeO  \cite{FeO}. The large incoherent
component of Fe 3$d$ at high binding energy is an indication of
relatively strong correlation effects in this material. The
correlation effects are also reflected from the observation that the
coherent part of Fe 3$d$ is compressed to 1 eV below $E_F$ (see
Fig.~1c) from the 2 eV range predicted by LDA. We note that this
narrowing effect has been predicted by a dynamic mean field theory
(DMFT) calculation which assumes large correlation effects
\cite{DMFT_Kotliar}.

Within the coherent Fe 3$d$ region, we observe several dispersive
bands, as shown in Fig.~2 which displays band dispersions along
several high symmetry directions ($\Gamma$-M, $\Gamma$-X, and M-X)
determined using the EDCs (Figs.~2a, b), $E$ $vs$ $k$ intensity
plots (Figs.~2c, d), and second derivative plot (Fig.~2e), and the
extracted EDC peak positions (Fig.~2f). We also plot the band
dispersion calculated from our LDA \cite{LDA_Xu} at this doping
level. The LDA bands, when normalized (divided) by a factor of 2,
agree well with the overall measured Fe 3$d$ bands, especially for
the high energy (0.2 - 0.6 eV) branches, as indicated in Fig.~2f.
This band narrowing factor of 2 , similar to an earlier ARPES
observation in a pnictide (LaFeOP) \cite{ShenP}, indicates the
importance of correlation effects as seen in the multiorbital
cobaltates \cite{cobaltate}. However, we observe
additional mass renormalization for the lower energy branches (below
0.2 eV) as shown in Fig.~2f and discussed in more details below,
suggesting a possible stronger and/or orbital (energy) dependent
correlation effect.

In order to accurately determine the low-energy band structure and
the FS, we have performed high-resolution ARPES measurements in the
vicinity of $E_F$, as shown in Fig. 3. Fig.~3a is an $E$ $vs$ $k$
intensity plot near the $\Gamma$ point, which clearly shows two
dispersing bands ($\alpha$ and $\beta$) forming two hole-like FS
pockets around $\Gamma$. This spectrum is measured in the
superconducting state where the quasiparticle (QP) peak width is
much narrower and the separation of the $\alpha$ and $\beta$ bands
is more visible. However, we did not observe a third band predicted
by LDA calculations, suggesting that it may be degenerated with
either the $\alpha$ or $\beta$ band. The narrow QP peak and high
resolution put an upper limit of 10 meV to the band splitting
between the two overlapping bands in the vicinity of $E_F$. We have
observed the top of the $\alpha$ band at the $\Gamma$ point by
measuring it at high temperature ($T$ = 150 K) and dividing the
spectrum by the Fermi function. As shown in Fig.~3b, this procedure
unmasks the spectrum within a few $k_BT$ above $E_F$, showing a
clear parabola with the band top situated at $\sim$ 20 meV above
$E_F$. This band top is much lower than the value of $\sim$ 120 meV
predicted by LDA calculations.

The band structure near M has been controversial. While LDA
predicted two electron-like Fermi surfaces around M, earlier ARPES
studies claimed to have observed a hole-like dispersion
\cite{ZhouBK,Borisenko}. Using high-resolution ARPES, in Figs.~3c
and d, we identify two electron-like bands (labeled as $\gamma$ and
$\delta$ bands) with the bottoms at $\sim$ 15 and 60 meV,
respectively. The energy difference between the observed $\alpha$
band top and $\gamma$ band bottom is about 35 meV, much smaller than
the value of $\sim$ 200 meV predicted by LDA \cite{LDA122_Lu}. In
addition, we observe in Fig.~3c a third band dispersing toward $E_F$
when moving from $\Gamma$ to M, intersecting with the $\delta$ band
in the vicinity of $E_F$. The anti-crossing of the two bands, or a
Dirac point, near M has been predicted by LDA, as can be seen in the
Fig.~4e. However, it occurs at a much
higher binding energy ($\sim$ 120 meV). This uplifting of the Dirac
point also creates a ``bright" spot with a strong intensity at $E_F$
observed in previous ARPES measurements
\cite{HongEPL,ZhouBK,Borisenko}.

We summarize our measurements of the FSs and the band structure in
Fig.~4. By measuring many cuts in the Brillouin zone (BZ), we have
obtained the FSs of Ba$_{0.6}$K$_{0.4}$Fe$_2$As$_2$. Fig.~4a
displays the ARPES intensity integrated within a narrow energy
window at $E_F$ ($\pm$10 meV), where the high intensity contours are
expected to follow the FS contours. We also extracted the $k_F$
points from the dispersive bands indicated by the dots in Fig.~4a
and obtained from both the momentum distribution curves (MDCs) and
the EDCs. These extracted $k_F$ points, in agreement with the
high-intensity contours, clearly show four FS sheets. The hole-like
$\alpha$ and $\beta$ FSs centered at $\Gamma$ enclose areas of 4\%
and 18\% of the BZ area, while the electron-like $\gamma$ and
$\delta$ FSs centered at M have areas of 2\% and 4\%, respectively.
According to Luttinger theorem on two-dimensional (2D) FS sheets,
the observed four FS sheets correspond to 16\% hole/Fe. The 40\% K
content in this optimally doped superconductor corresponds to 20\%
hole/Fe, suggesting that the fifth FS sheet predicted by LDA would
be degenerated with the $\alpha$ FS, yielding a total Luttinger area
of 20\% hole/Fe.

The above analysis of Luttinger area is under the assumption of 2D
FS. However, LDA calculations predict significant $k_z$ dispersions
\cite{LDA122_Lu}. To check this, we have measured the $\alpha$ and
$\beta$ FSs using different photon energies, or equivalently, at
different values of $k_z$. The results are shown in Fig.~4b where
the extracted FS areas are plotted as functions of $k_z$, which are
estimated by using the free-electron final state approximation with
a 10 eV inner potential. From Fig.~4b we do not observe a strong
periodic variation of the FS area as $k_z$ is changed, in contrast
to the LDA prediction of strong $k_z$-dependent FS sheets. The
quasi-2D FSs are thought to be at odds with the isotropic upper
critical field ($H_{c2}$) derived from a high-field measurement
\cite {Yuan_Hc2}. However, a recent sensitive magnetic torque
experiment on high-quality single crystals has observed a larger
anisotropy between the $a$-$b$ plane and the $c$-axis \cite{torque}.

We plot our measurements of band dispersion along the two
high-symmetry lines ($\Gamma$-M and $\Gamma$-X) in Fig.~4e. As a
comparison, we also plot the LDA bands renormalized by a factor of
2. While there is a fairly good agreement on the Fermi crossing
($k_F$) between the ARPES measurements and LDA calculations, the
Fermi velocity ($v_F$) is further renormalized as compared to the
renormalized LDA bands. In the table of Fig.~4c, we list the values
of $v_F$ along $\Gamma$-M obtained from ARPES measurement and LDA
calculations. Most low energy bands have a mass renormalization greater than 4,
significantly larger than the overall band renormalization obtained
at high binding energy. This additional renormalization at low
energy, which has also been observed in both cuprates and cobaltates
\cite{Valla,cobaltate}, can be attributed to an enhanced self-energy
effect near $E_F$. Possible causes of the enhanced self-energy near
$E_F$ include electronic correlation effects, low-energy modes or
fluctuations, and band hybridization. Independent of its microscopic
origin, the small renormalized Fermi velocity has an important
consequence on the superconducting coherence length, which can be
estimated from the BCS relation $\xi = \frac{ \hbar
v_F}{\pi\Delta}$. From the obtained values of $v_F$ and the measured
superconducting gaps \cite{HongEPL}, we estimate that the coherence
length in this superconductor is about 9 - 14 \AA, which agrees
well with the value ($\sim$ 12 \AA) obtained from the $c$-axis upper
critical field \cite{SH_Kwok}. 
It is remarkable that the superconducting coherence length
in this pnictide is much smaller than conventional BCS
superconductors, and surprisingly close to that found in the cuprate
superconductors.

Basic band parameters of this material constitute important starting
point for microscopic and phenomenological theories of its unusual
superconductivity and other instabilities. After determining the
band structure and the FSs, we can now extract the dispersion
parameters. Here we adapt a simple tight-binding-like band structure
proposed previously \cite{Eremin}: $E^{\alpha,\beta}(k_x,k_y) =
E_0^{\alpha,\beta} + t_1^{\alpha,\beta}(\cos{k_x}+\cos{k_y}) +
t_2^{\alpha,\beta}\cos{k_x}\cos{k_y}$, and
$E^{\gamma,\delta}(k_x,k_y) = E_0^{\gamma,\delta} +
t_1^{\gamma,\delta}(\cos{k_x}+\cos{k_y}) +
t_2^{\gamma,\delta}\cos{(k_x/2)}\cos{(k_y/2)}$. Under mild
constraints for the unoccupied bands which are not accessible by
ARPES, we obtain the band parameters for the four observed bands
which are listed in the table in Fig.~4e. As shown in Figs.~4d and
e, the FS contours and the band structure generated from these
parameters match remarkably well with the observed FSs and band
dispersions. Note that the fitted FSs for the electron-like bands in
the vicinity of M are two ellipses without including the effects of
band hybridization between them. The latter splits the two bands and
give rise to the observed outer ($\delta$) and inner ($\gamma$) FSs.
As seen clearly in Fig.~4d, the $\alpha$ FS, when shifted by the
($\pi$, $\pi$) wavevector, overlaps well with the $\delta$ FS,
strongly suggesting a good FS nesting with the same nesting
wavevector ($\pi$, $\pi$) as the AF wavevector observed in the
parent compound \cite {neutron_Dai}. Combined with the observation
of nearly identical superconducting gaps on the $\alpha$,  $\delta$,
and $\gamma$ FSs \cite{HongEPL,4thgap} in the same material, this
result reinforces the notion that the interband scattering among the
nested Fermi surfaces plays a dominant role in the pairing
interaction \cite{Mazin,Kuroki,Lee,Hu}.

One application for these band parameters is the estimate of the
effective masses of the low energy bands, whose average values along
the FSs are obtained: $m^{\ast}_{\alpha} = 4.8$, $m^{\ast}_{\beta} =
9.0$, $m^{\ast}_{\gamma} = 1.3$, $m^{\ast}_{\delta} = 1.3$, in units
of free electron mass. From these effective masses, we deduce the
Sommerfeld parameter of the specific heat by using $\gamma = \pi
N_AK^2_Ba^2m^{\ast}/3\hbar^2$ (where $a$ = 3.92 \AA \xspace is the
lattice parameter and $N_A$ is Avogadro's number), which are 7.2,
13.6, 2, 2 in units of mJ/K$^2$mol for the $\alpha$, $\beta$,
$\gamma$, $\delta$ bands, respectively. The total value of the
$\gamma$ coefficient, assuming doubly degenerated $\alpha$ bands, is
$\sim$ 32 mJ/K$^2$mol, which is much larger than the value ($\sim$ 9
mJ/K$^2$mol) calculated by LDA \cite{LDA122_Lu}, reflecting the
effect of strong mass renormalization. The experimentally obtained
$\gamma$ coefficient in the parent compound SrFe$_2$As$_2$ is 6.5
mJ/K$^2$mol \cite{ChenGF}, which is expected to be significantly
reduced since large portions of FS sheets are gapped
\cite{optical,quantum}. The normal state $\gamma$ coefficient in the
superconducting Ba$_{0.6}$K$_{0.4}$Fe$_2$As$_2$, although difficult
to obtain due to its high critical field ($H_{c2}$), was estimated
to be as large as $\sim$ 63 mJ/K$^2$mol \cite{SH_Mu}.

In conclusion, we have fully determined the band structure and Fermi
surfaces of the optimally doped Fe-based superconductor
Ba$_{0.6}$K$_{0.4}$Fe$_2$As$_2$. Four dispersive bands in the
vicinity of $E_F$ have been clearly observed, resulting in two
hole-like FSs (with a possible third degenerated FS) around $\Gamma$
and two electron-like FSs around M. The two electron-like FSs are
consistent with the unexpected hybridization of the two overlapping
orthogonal elliptic FSs, resulting in the outer ($\delta$) FS being
nearly nested with the inner hole-like ($\alpha$) FS via the ($\pi$,
$\pi$) AF wavevector. In contrast to the LDA, only small $k_z$
dispersions are observed for the FSs, with the total volume
consistent with Luttinger theorem. Furthermore, the overall
bandwidth is renormalized by a factor of 2, while the low energy
dispersions acquire an even stronger renormalization effect that
implies a much enhanced specific heat coefficient $\gamma \sim$ 32
mJ/K$^2$mol calculated using the tight-binding band parameters
obtained from fitting the observed electronic structure. The
comparison of our measured band dispersions with those predicted by
LDA provides valuable insights into the possible effects of
electronic correlations, a central unresolved issue in the field of
iron pnictides.

This work was supported by grants from Chinese Academy of Sciences,
NSF, Ministry of Science and Technology of China, JSPS, TRIP-JST,
CREST-JST, MEXT of Japan, and NSF (DMR-0800641, DMR-0704545), DOE
(DEFG02-99ER45747) of the US.  This work is based upon research
conducted at the Synchrotron Radiation Center supported by NSF
DMR-0537588, and the Advanced Light Source supported by DOE No.
DE-AC02-05CH11231.

\bibliography{biblio_en}

\begin{thebibliography}{10}
\bibitem{footnote} We caution that some ordinate Auger intensity will be also included in the difference curve.
\end{thebibliography}

\bigskip

\begin{figure}[htbp]
\begin{center}
\includegraphics[width=5in]{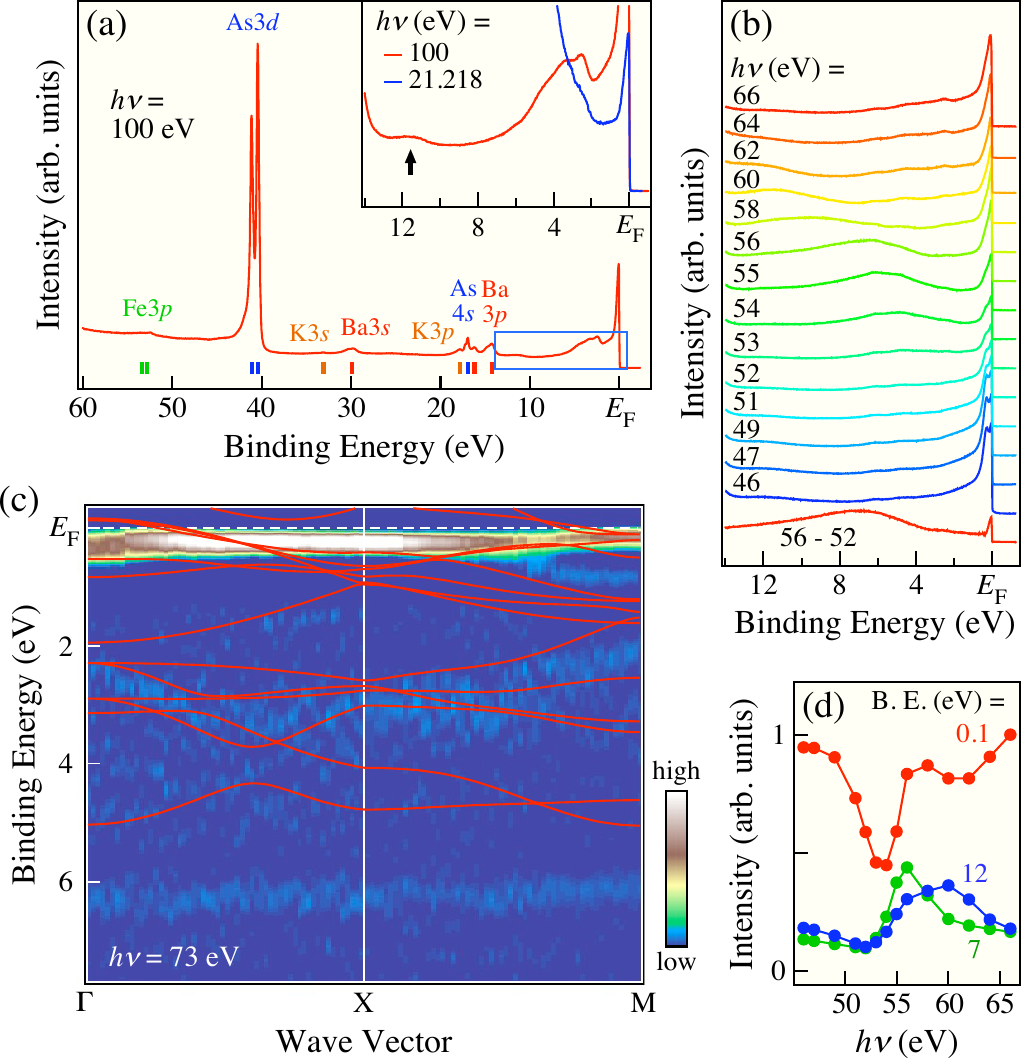}
\caption{\label{Fig1}
Shallow core levels and valence band of Ba$_{0.6}$K$_{0.4}$Fe$_2$As$_2$. (a) Wide range EDC near $\Gamma$ ($h\nu$ = 100 eV) showing many shallow core levels, whose atomic energy levels are marked by vertical bars above the $x$-axis.  The inset magnifies the valence band and a possible satellite peak at $\sim$ 12 eV, and highlights the difference between spectra taken at 100 and 21.2 eV. (b) Valence band near $\Gamma$ measured at different photon energies (46 - 66 eV). The red spectrum at the bottom is the difference
between EDCs measured at 56 (at resonance) and 52 eV (below resonance). All EDCs are normalized by the photon flux.  (c) Intensity plot of second derivatives of spectra along $\Gamma$-X and X-M ($h\nu$ = 73 eV). LDA bands (red lines) are also plotted for comparison. (d) Photon energy dependence of the intensity of
EDCs shown in Fig.~1b, obtained at the binding energies of 0.1, 7, and 12 eV . }
\end{center}
\end{figure}

\begin{figure}[htbp]
\begin{center}
\includegraphics[width=5in]{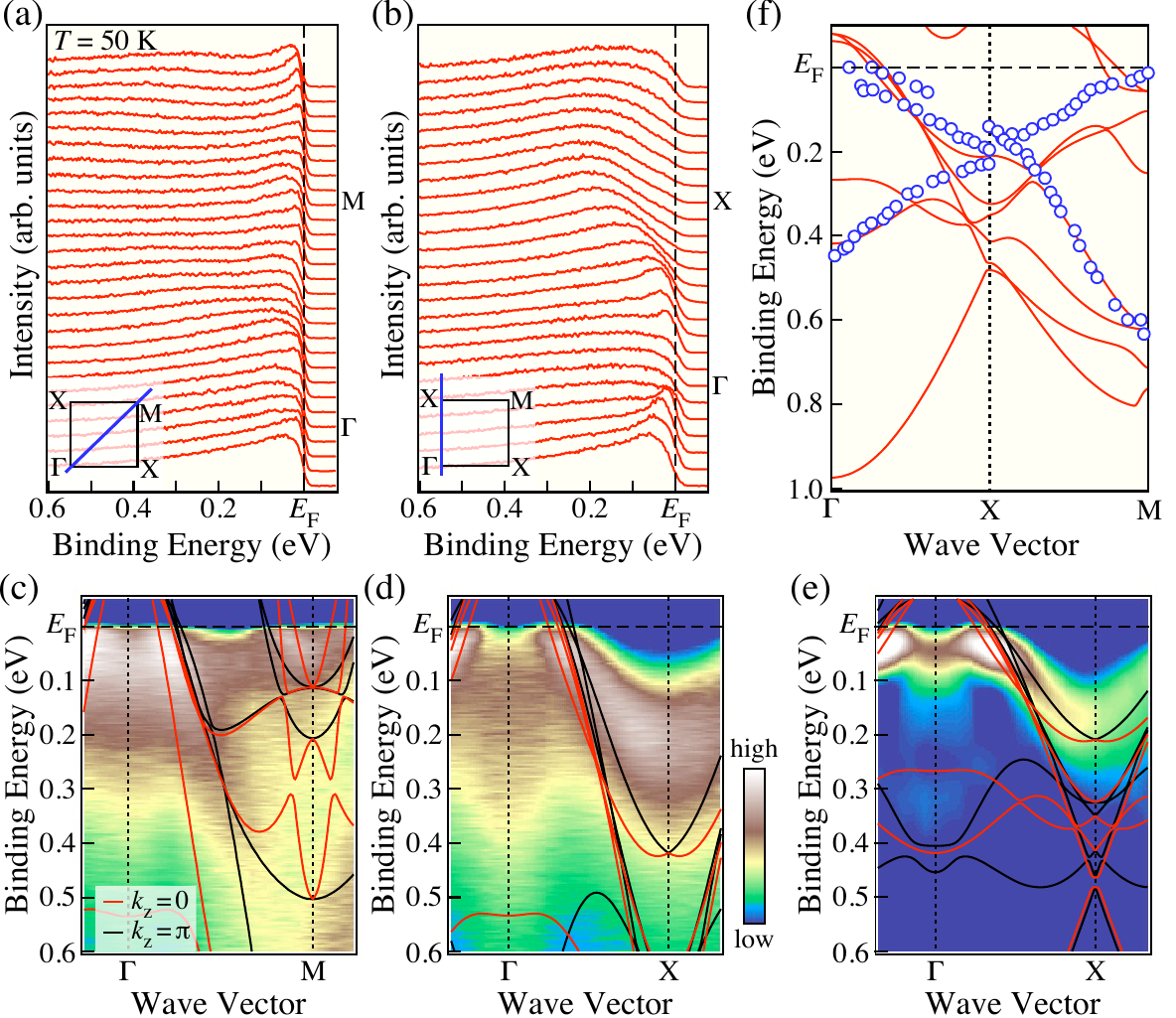}
\caption{\label{Fig2}
Coherent Fe 3$d$ spectra within 0.6 eV from $E_F$ measured at 50 K along $\Gamma$-M and $\Gamma$-X.
EDCs along (a) $\Gamma$-M and (b) $\Gamma$-X ($h\nu$ = 21.2 eV). Intensity plots of the same spectra along (c)
$\Gamma$-M and (d) $\Gamma$-X. LDA calculated bands at $k_z$ = 0 (red) and $k_z$ = $\pi$ (black) are also plotted
for comparison. (e) Intensity plot of second derivatives of the spectra along $\Gamma$-X in comparison with the
LDA bands renormalized by a factor of 2. (f) Extracted band positions (blue circles) measured at 45 eV along $\Gamma$-X-M with the
comparison to the same renormalized LDA bands of $k_z$ = 0.
}
\end{center}
\end{figure}

\begin{figure}[htbp]
\begin{center}
\includegraphics[width=5in]{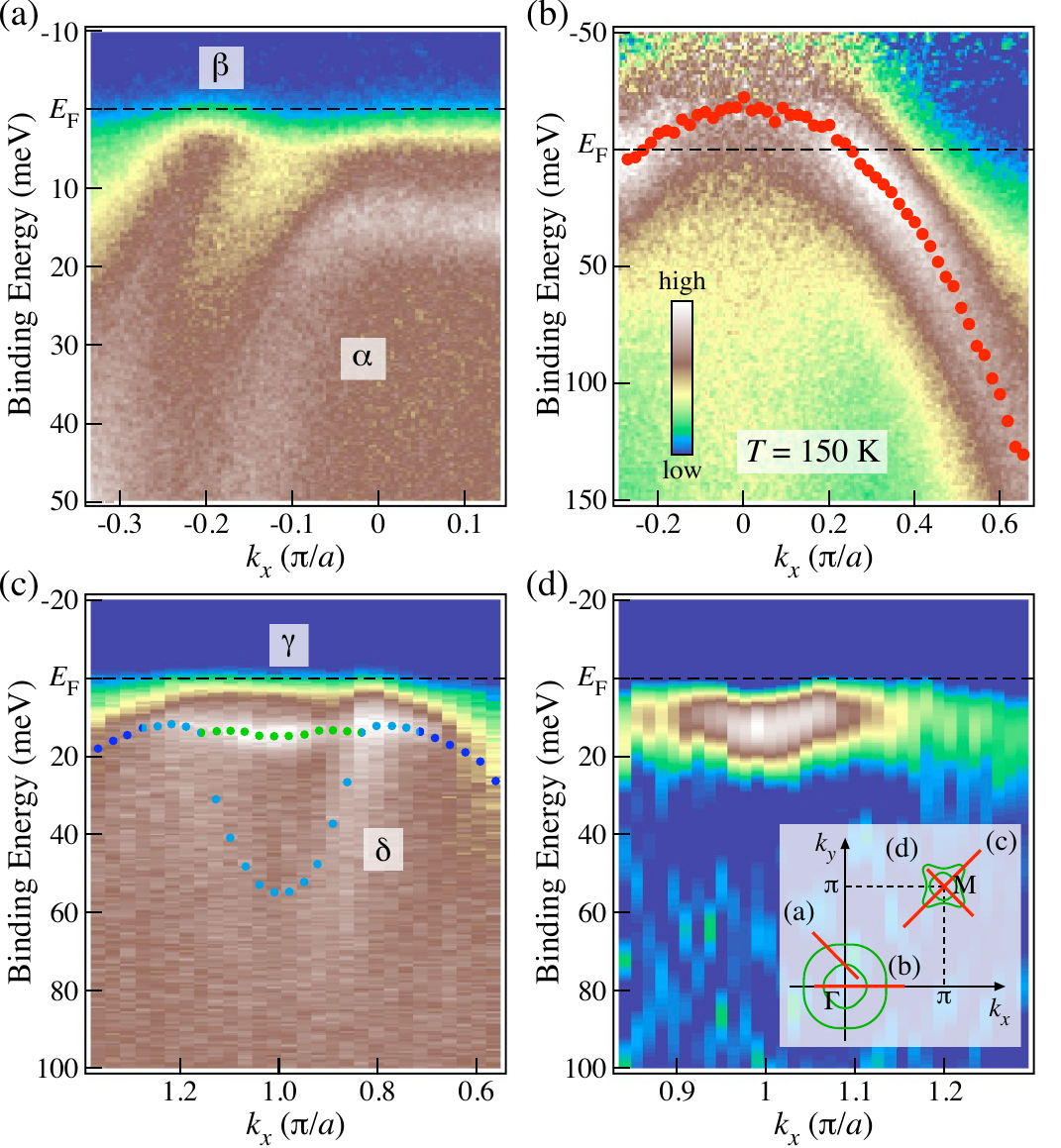}
\caption{\label{Fig3}
Low-energy fine structures of Ba$_{0.6}$K$_{0.4}$Fe$_2$As$_2$.
(a) Intensity plot near $\Gamma$ measured at low temperature ($T$ = 15 K) clearly displaying two hole-like ($\alpha$ and $\beta$) bands approaching $E_F$.
(b) Intensity plot near $\Gamma$ measured at high temperature ($T$ = 150 K) divided by the Fermi function, showing that the band top of the $\alpha$ band is
$\sim$ 20 meV above $E_F$. The red dots are band positions extracted from EDC peaks. (c) Intensity plot in the vicinity of M measured at 15 K showing two
electron-like ($\gamma$ and $\delta$) bands and a hole-like band dispersion approaching $E_F$ (see blue dots). The dots are EDC peak positions.
(d) Intensity plot of second derivatives of the spectra near M, indicating the electron-like nature of the $\gamma$ band. The inset indicates measurement locations
in the BZ for panels (a) - (d).
}
\end{center}
\end{figure}

\begin{figure}[htbp]
\begin{center}
\includegraphics[width=5in]{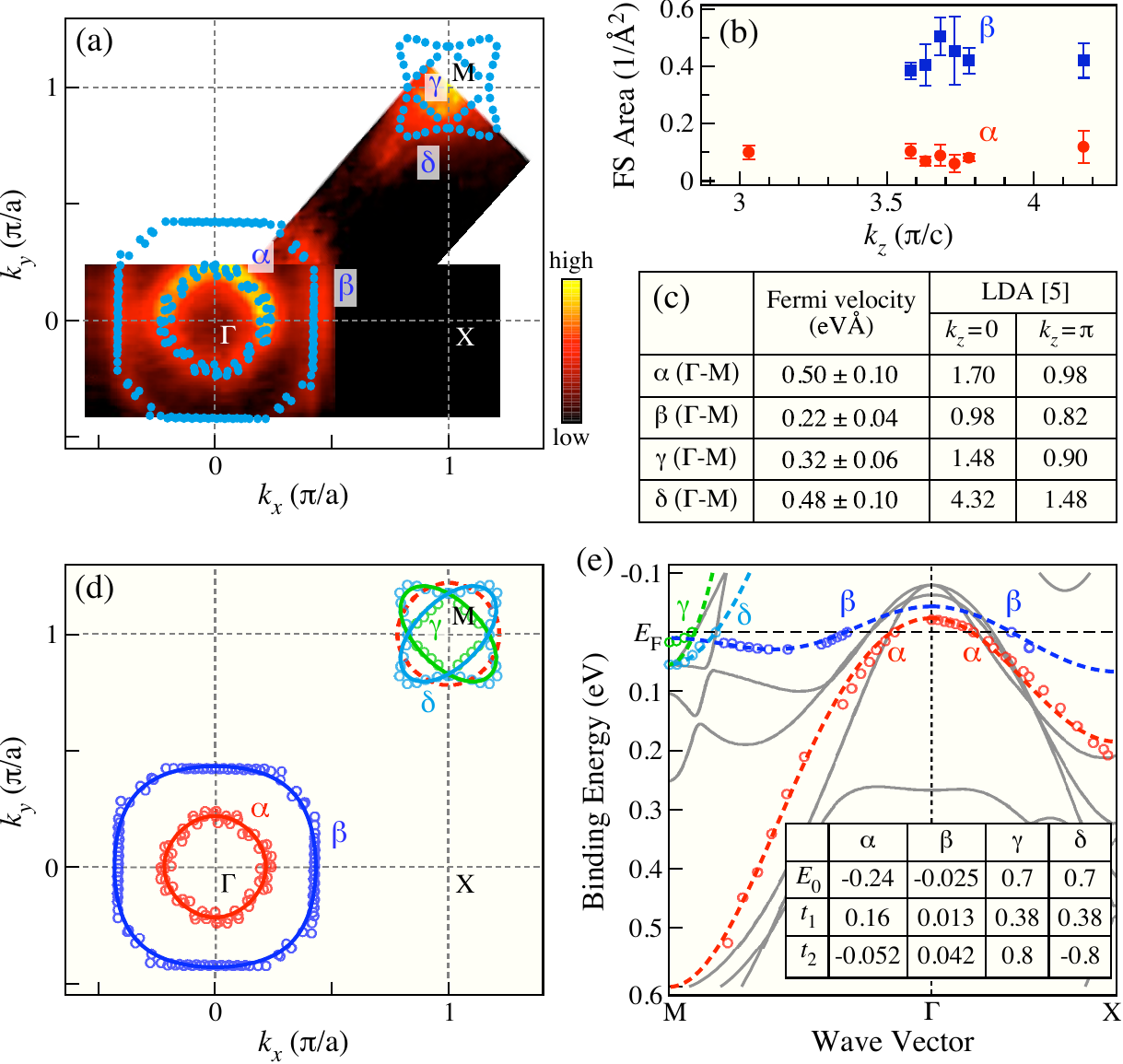}
\caption{\label{Fig4}
Summary of measured FSs and band structure.
(a) FSs on the 2D BZ obtained from the intensity integrated within $E_F\pm$10 meV (false color plot) and the Fermi crossing points (blue dots)
obtained from MDC and EDC dispersions. (b) $k_z$ dispersion of the $\alpha$ and $\beta$ FS areas obtained by changing photon energy.
(c) Table of the measured and calculated Fermi velocities along principle axes.
(d) Measured FS (circles) and tight-binding fitting curves (solid lines). The dashed line is the fitted $\alpha$ FS shifted by the ($\pi$, $\pi$) wavevector.
(e) Measured band dispersion (circles) along $\Gamma$-M and $\Gamma$-X, compared with the LDA bands renormalized by a factor of 2 (solid lines),
and tight-binding fits (dashed lines). The inserted table lists the parameters of tight-binding bands.
}
\end{center}
\end{figure}

\end{document}